\documentclass[useAMS,usenatbib]{mn2e}
\usepackage{natbib}
\usepackage{graphicx}\usepackage{epsfig}\usepackage{epsf}
\usepackage{amsmath}\usepackage{amssymb}\usepackage{stfloats}
\title[NGC~4490 OT]{Massive-Star Mergers and the Recent Transient in
  NGC4490: A More Massive Cousin of V838~Mon and V1309~Sco}

\author[Smith et al.]{Nathan Smith$^{1}$\thanks{E-mail:
    nathans@as.arizona.edu}, Jennifer E.\ Andrews$^1$, Schuyler D.\
  Van Dyk$^2$, Jon C.\ Mauerhan$^{3}$, \newauthor Mansi M.\
  Kasliwal$^4$, Howard E.\ Bond$^{5,6}$, Alexei V. Filippenko$^3$,
  Kelsey I.\ Clubb$^3$, \newauthor Melissa L.\ Graham$^2$, Daniel
  A. Perley$^{4,7}$, Jacob Jencson$^{4,8}$, John Bally$^9$, \newauthor Leonardo
  Ubeda$^6$, Elena Sabbi$^6$ \\
  $^{1}$Steward Observatory, University of Arizona, 933 N. Cherry Ave., Tucson, AZ 85721, USA \\
  $^2$Spitzer Science Center/Caltech, Mail Code 220-6, Pasadena, CA 91125, USA \\
  $^3$Department of Astronomy, University of California, Berkeley, CA
  94720-3411, USA \\ $^4$Astronomy Department, California Institute of
  Technology, 1200 E. California Boulevard, Pasadena, CA 91125, USA \\
  $^5$Department of Astronomy \& Astrophysics, Pennsylvania State
  University, University Park, PA 16802, USA \\
  $^6$Space Telescope Science Institute, 3700 San Martin Dr., Baltimore, MD 21218, USA \\
  $^7$Dark Cosmology Centre, Niels Bohr Institute, University of Copenhagen, Denmark \\
  $^8$NSF Graduate Fellow \\ $^9$Center for Astrophysics and Space
  Astronomy, University of Colorado, 389 UCB, Boulder, CO 80309, USA}
\begin{document}
\pagerange{\pageref{firstpage}--\pageref{lastpage}} \pubyear{2015}
\maketitle
\label{firstpage}

\begin{abstract}

  The Galactic transient V1309~Sco was the result of a merger in a
  low-mass star system, while V838~Mon was thought to be a similar
  merger event from a more massive B-type progenitor.  In this paper
  we study a recent optical and infrared (IR) transient discovered in
  the nearby galaxy NGC~4490 named NGC~4490-OT2011 (NGC~4490-OT
  hereafter), which appeared similar to these merger events
  (unobscured progenitor, irregular multi-peaked light curve,
  increasingly red colour, similar optical spectrum, IR excess at late
  times), but which had a higher peak luminosity and longer duration
  in outburst. NGC~4490-OT has less in common with the class of
  SN~2008S-like transients.  A progenitor detected in pre-eruption
  {\it Hubble Space Telescope} ({\it HST}) images, combined with upper
  limits in the IR, requires a luminous and blue progenitor that has
  faded in late-time {\it HST} images.  The same source was detected
  by {\it Spitzer} and ground-based data as a luminous IR (2--5
  $\mu$m) transient, indicating a transition to a self-obscured state
  qualitatively similar to the evolution seen in other stellar mergers
  and in luminous blue variables.  The post-outburst dust-obscured
  source is too luminous and too warm at late times to be explained
  with an IR echo, suggesting that the object survived the event.  The
  luminosity of the enshrouded IR source is similar to that of the
  progenitor.  Compared to proposed merger events, the more massive
  progenitor of NGC~4490-OT seems to extend a correlation between
  stellar mass and peak luminosity, and may suggest that both of these
  correlate with duration.  We show that spectra of NGC~4490-OT and
  V838~Mon also resemble light-echo spectra of $\eta$ Car, prompting
  us to speculate that $\eta$ Car may be an extreme extension of this
  phenomenon.

\end{abstract}


\section{INTRODUCTION}

Ongoing dedicated studies of the transient sky are revealing an
increasingly wide diversity of explosive and eruptive stellar
transients that are less luminous than traditional supernovae (SNe),
but more luminous than classical novae.  There is a broad range of
theoretical mechanisms that might account for these, yet linking
physical mechanisms to individual observed phenomena remains
challenging.  Extragalactic events being discovered in modern
transient surveys occur at distances where detailed information about
the progenitor star or potential surviving star is scarce,
complicating any connection to observed Galactic populations.

More than a decade ago, few such events were known, and the transients
with luminosities between those of novae and SNe were generally linked
to the class of eruptive massive stars known as luminous blue
variables (LBVs; \citealt{hd94}).  Names like SN impostors, Type~V
SNe, or $\eta$ Carinae variables were often taken to be synonyms of
LBVs \citep{hd94,hds99,svd00}.  With the increasing observed diversity
of the SN impostor-like transients, however, some differentiation in
observed properties also developed.  Proposed observational subclasses
include LBV giant eruptions \citep{smith11}, SN~2008S-like events
\citep{thompson09,prieto+08,kochanek11}, and a class of luminous red
variables that have been linked to collisions and merger events (see
below).

The SN~2008S-like sources, including the well-studied event
NGC~300-OT2008 \citep{bond09} and its kin, seem to be distinguished in
that they have heavily dust-enshrouded progenitors, more monotonic
optical light-curve evolution, strong [Ca~{\sc ii}] and Ca~{\sc ii}
emission features, and strong infrared (IR) excess at late times
\citep{thompson09,prieto+08,kochanek11}.  Some LBVs share these
properties \citep{smith11}, but the SN~2008S-like progenitors tend to
have lower luminosities and implied masses
\citep{thompson09,prieto+08,kochanek11} than the recognized mass and
luminosity ranges for LBVs \citep{smith04}.  In addition, unlike LBVs,
there is no clear sign that SN~2008S-like transients survive the event
\citep{adams15}.  The class of red transients identified as possible
mergers also tend to have lower-mass progenitors when this information
is available, but their progenitors do not appear to be as heavily
enshrouded in dust as the SN~2008S-like objects. Moreover, the
SN~2008S-like objects and the lower-mass merger outbursts seem to
trace different Galactic stellar populations.  In this paper, we
suggest a new potential member of this class of merger objects, but
with a higher initial mass than any object in this class studied so
far.  This new object in NGC~4490 also shares many properties in
common with giant LBV eruptions, so the distinction between LBVs and
merger events may be blurry.


\citet{tylenda11} presented the most direct and dramatic evidence
linking one of these transients to a violent stellar merger event.
They analyzed archival Optical Gravitational Lensing Experiment (OGLE)
photometry and found that the progenitor of the 2008 Galactic
transient V1309~Sco had been an eclipsing binary, and that the orbital
period decreased rapidly leading up to the catastrophic transient
event.  The progenitor star was thought to have an initial mass of
1--2 M$_{\odot}$, while the transient event had a peak absolute
magnitude of roughly $-$6.9 mag ($V$ band; Vega-based magnitudes),
although the distance is uncertain, and a duration of about a month.
A brighter Galactic event was the 2002 transient V838~Mon, with a peak
absolute visual magnitude of about $-$9.8, a longer duration of about
80--90 d, and a very irregular, multi-peaked light curve
\citep{ab07,bond03,sparks08}.  The progenitor of this transient was
thought to be more massive (5--10 M$_{\odot}$), due to a small host
cluster with B-type stars still on the main sequence
\citep{rr08,ab07}.  It also had a B3~V companion star that was
engulfed by the expanding dusty ejecta several years after the
outburst. V838~Mon is perhaps best known for the series of spectacular
colour light-echo images taken by the {\it Hubble Space Telescope}
({\it HST}) \citep{bond03}.

Both V1309~Sco and V838~Mon started out with relatively smooth
continuum emission plus narrow Balmer emission lines, but developed
very red optical colours, atomic and molecular absorption features,
and strong signatures of dust in their spectra as they faded
\citep{munari07,mason10,nichols13,loebman15}.  The observed properties
of these transients appear to be consistent with theoretical
expectations for merger events \citep{ivanova13,st06,st07}.  Other
events that may belong to the same class are the Galactic transients
V4332 Sgr in 1994 \citep{martini99} and OGLE 2002-BLG-360
\citep{tylenda13}, and potentially also the historical Galactic novae
CK~Vul \citep{kato03,km15}, and V1148~Sgr \citep{mayall49}.  Proposed
extragalactic counterparts include M31~RV \citep{rich89,bs06,bond11},
M85~OT2006-1 \citep{srk07}, and the transient in M31 discovered in
January 2015 \citep{kurtenkov15,williams15}.  \citet{kochanek14} have
estimated the Galactic rates for merger events and found them to be
quite common, with approximate values of 0.1 yr$^{-1}$ for fainter
events like V1309~Sco, and 0.03 yr$^{-1}$ for brighter events like
V838~Mon. \citet{kochanek14} also suggest that the peak luminosity of
the resulting transient is a steep function of the stellar mass,
similar to the mass-luminosity relation on the main sequence.


In this paper we discuss the bright optical and IR transient source
that appeared in 2011 in the host galaxy NGC~4490, and we demonstrate
that it bears some interesting similarities to the class of mergers
discussed above.  We also point out that despite its similar peak
luminosity, this new event appears to have less in common with
SN~2008S-like transients. The object was first recognized as a
transient in the optical, so for brevity we refer to it here as
NGC~4490-OT, although subsequently it was also shown to be a transient
source in the thermal IR. It is also designated as
PSN~J12304185+4137498, and in our IR transient search described below
it was found independently and named SPIRITS~14pz.

\begin{figure*}
\includegraphics[width=6.5in]{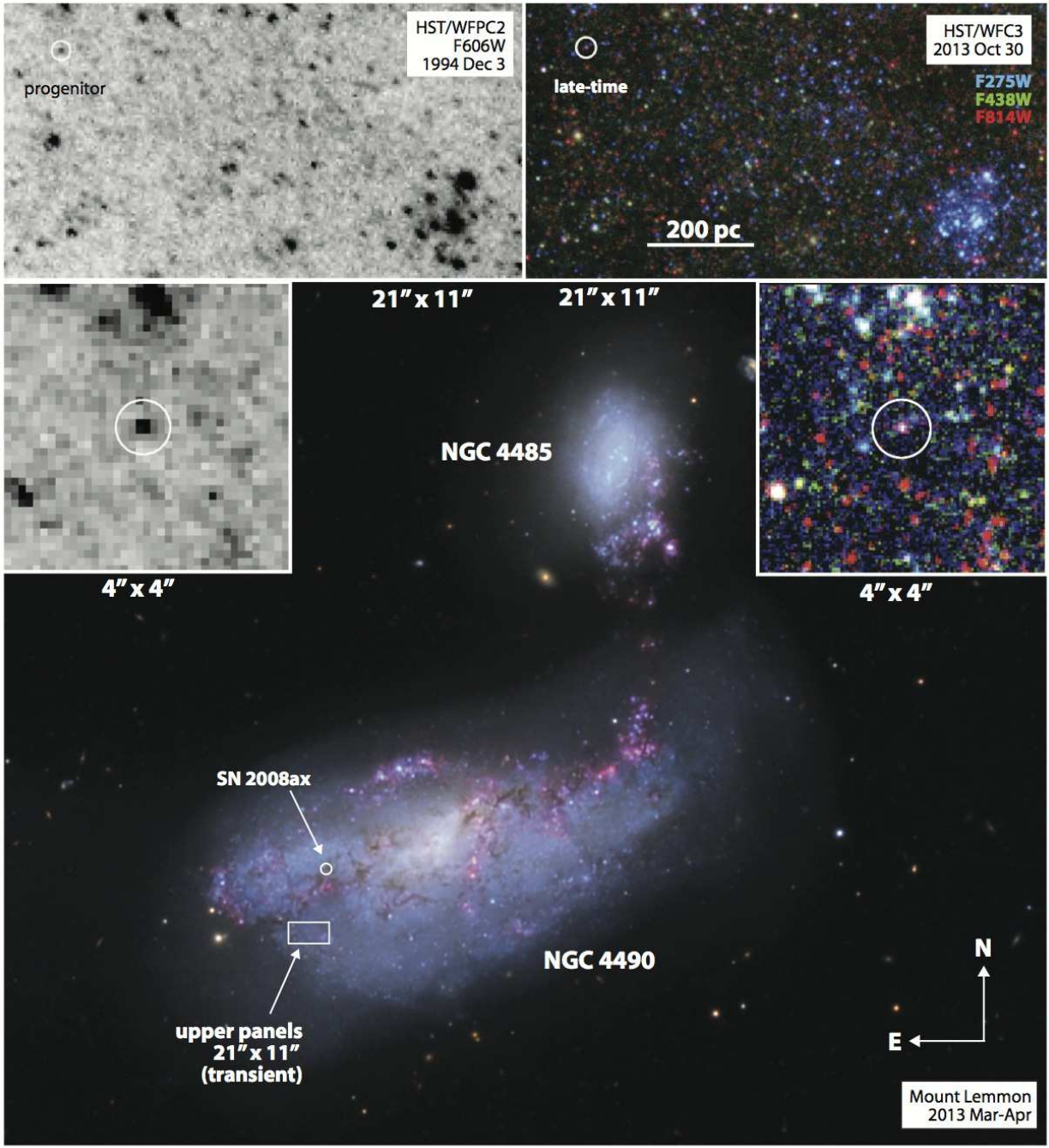}
\caption{Images of the host environment of the NGC~4490 transient.
  The main image at the bottom shows a wide-field view of the
  interacting pair of galaxies NGC~4490 and NGC~4485.  This colour
  image is a composite of individual visible-wavelength images taken
  on different nights during March and April 2013 using the 0.8~m
  Schulman Telescope at the Mount Lemmon SkyCenter (image credit: Adam
  Block/Mount Lemmon SkyCenter/University of Arizona; used with
  permission).  This image is included to show the location and
  surrounding environment of the transient within the host galaxy.
  The small white rectangle in the lower left marks the immediate
  environment of the transient, corresponding to the 21{\arcsec}
  $\times$ 11{\arcsec} inset images at the top of the figure.  The
  position of SN~2008ax is also noted.  These insets show {\it
    HST} images, with the grey-scale image at left corresponding to the
  pre-eruption archival image taken with WFPC2 in the F606W filter in
  1994.  The inset at right is a colour image made from F275W, F438W,
  and F814W frames obtained with the WFC3-UVIS camera after the
  eruption had faded in October 2013.  In these images, the white
  circle marks the position of the NGC~4490-OT transient.  The
  4{\arcsec}$\times$4{\arcsec} boxes are zoomed-in sections of these
  same two {\it HST} frames.}
\label{fig:hst}
\end{figure*}

NGC~4490-OT was discovered in unfiltered CCD images on 2011 Aug 16.83
(UT dates are used throughout this paper) by \citet{ca11}, located
57{\arcsec} west and 37{\arcsec} south of the bright centre of the
nearby galaxy NGC~4490.  Its coordinates are $\alpha_{2000}$ =
12$^h$30$^m$41$\fs$84; $\delta_{2000}$ =
+41$^{\circ}$37$\arcmin$49$\farcs$7.  NGC~4490 is part of a pair of
actively star-forming interacting galaxies (the other galaxy in the
pair is NGC~4485; see Figure~\ref{fig:hst}), and NGC~4490 was also the
host galaxy as for the Type~IIb event SN~2008ax.  According to the
NASA Extragalactic Database
(NED)\footnote{https://ned.ipac.caltech.edu/}, the host has an average
redshift of $z=0.00185$ (565 km s$^{-1}$).  This would imply a
distance of $d=9.22$ Mpc and a distance modulus of $m-M=29.82$ mag.
However, in their study of SN~2008ax, \citet{pastorello08} noted some
inconsistencies in the distances to the two galaxies in the
interacting pair and derived an average distance combining several
techniques of $d=9.6$ Mpc and $m-M=29.92$ mag.  We adopt these latter
values here.  We also adopt a line-of-sight Milky Way reddening of
$E(B-V)=0.019$ mag ($A_V=0.06$ mag, $A_R=0.047$ mag) from
\citet{sf11dust}.  There may be additional host-galaxy reddening
(\citealt{pastorello08} estimate $E(B-V)=0.3$ mag for SN~2008ax, which
is in a similar part of the galaxy), or circumstellar reddening, but
we consider these in more detail later.

With the above adopted parameters, the discovery brightness of the
transient corresponds to an absolute magnitude (assumed to be roughly
$R$ band) of about $-$13.3 mag. As discussed below, the peak
luminosity 100--200 d later is somewhat higher, around $-$14.2 mag
(with the additional host-galaxy reddening that we favor in the
discussion below, the peak would be about $-$15 mag).  This implies
that the object may be related to the broad class of SN impostors that
are less luminous than core-collapse SNe.  Early classification
spectra obtained on 2011 Aug. 18.9 and 19.9 (days 2--3 after
discovery) showed a blue continuum and bright, narrow Balmer emission
lines, plus some fainter Fe~{\sc ii} lines \citep{magill11},
qualitatively similar to known spectra of LBV eruptions
\citep{smith11}.  However, \citet{magill11} noted that there was no
bright, blue star known at that position in previously obtained
archival {\it HST} images, while archival {\it Spitzer} data did not
reveal a bright mid-IR source either, apparently disfavouring a very
luminous LBV progenitor.

\citet{fraser11} conducted a more detailed analysis of the archival
{\it HST} images, and proposed that a source at the nominal position
of the transient was a likely progenitor detection, with an absolute
magnitude $-$6.2 in the F606W filter in the mid 1990s.  (With the
somewhat different distance and reddening we adopt, this corresponds
to $-$6.4 mag.)  This is much brighter at visual wavelengths than the
heavily dust-enshrouded progenitors of SN~2008S and NGC~300-OT
\citep{prieto08,prieto+08,thompson09}, and only modestly fainter than
the $-$7.5 mag progenitor of the LBV-like transient UGC~2773-OT
\citep{smith+10,smith16} and the LBV progenitor of the SN~1954J
eruption \citep{ts68,smith01,svd05}.  This may suggest that the
progenitor and its outburst are related to the LBV phenomenon in more
massive stars, but as we detail below, it also shares some interesting
properties with lower-luminosity transients that are thought to arise
from stellar merger events.

\begin{table}\begin{center}\begin{minipage}{3.25in}
      \caption{{\it HST} Imaging and photometry of NGC~4490-OT}
\scriptsize
\begin{tabular}{@{}llcccc}\hline\hline
Date     &Instr &Day &Filter &mag$^a$ &$\sigma$ \\ \hline
%
%
%
1994\,Dec\,03 &WFPC2  &$-$6100  &F606W  &23.58  &0.24   \\
2013\,Oct\,30 &WFC3-UVIS  &805  &F275W  &23.68  &0.07   \\
2013\,Oct\,30 &WFC3-UVIS  &805  &F336W  &25.36  &0.23   \\ 
2013\,Oct\,30 &WFC3-UVIS  &805  &F438W  &25.43  &0.10   \\
2013\,Oct\,30 &WFC3-UVIS  &805  &F555W  &25.33  &0.05   \\
2013\,Oct\,30 &WFC3-UVIS  &805  &F814W  &23.40  &0.04   \\
\hline
\end{tabular}\label{tab:hstphot}\end{minipage}
\end{center}
$^a$ The 1994 magnitude was measured in a 0$\farcs$5 diameter aperture.
The 2013 measurements are the result of PSF-fitting photometry from
the LEGUS photometry catalog (see text).  We also performed aperture photometry
in a small 0$\farcs$5 diameter aperture that yielded slightly
different magnitudes, as discussed later.
\end{table}

\begin{figure*}
\includegraphics[width=5.9in]{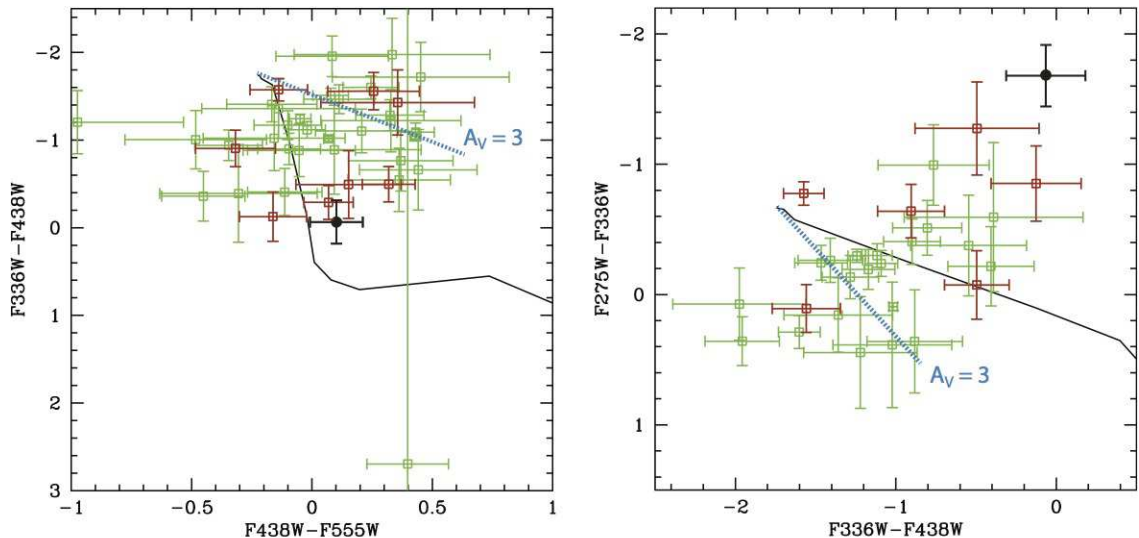}
\caption{Colour-colour plots from {\it HST} LEGUS data taken at late
  times, produced as described in the text, Section 2.1.  Green open
  squares correspond to detected sources within a projected separation
  of 100 pc from NGC~4490-OT, while red open squares are projected
  within 50 pc.  The filled black circle is NGC~4490-OT at late times.
  The colour-colour locus for normal supergiants is shown by the black
  curve in both plots, and a reddening vector for $A_V=3$ mag and
  $R=3.1$ is shown by the blue dashed line.}
\label{fig:cc}
\end{figure*}

\begin{figure*}
\includegraphics[width=6.3in]{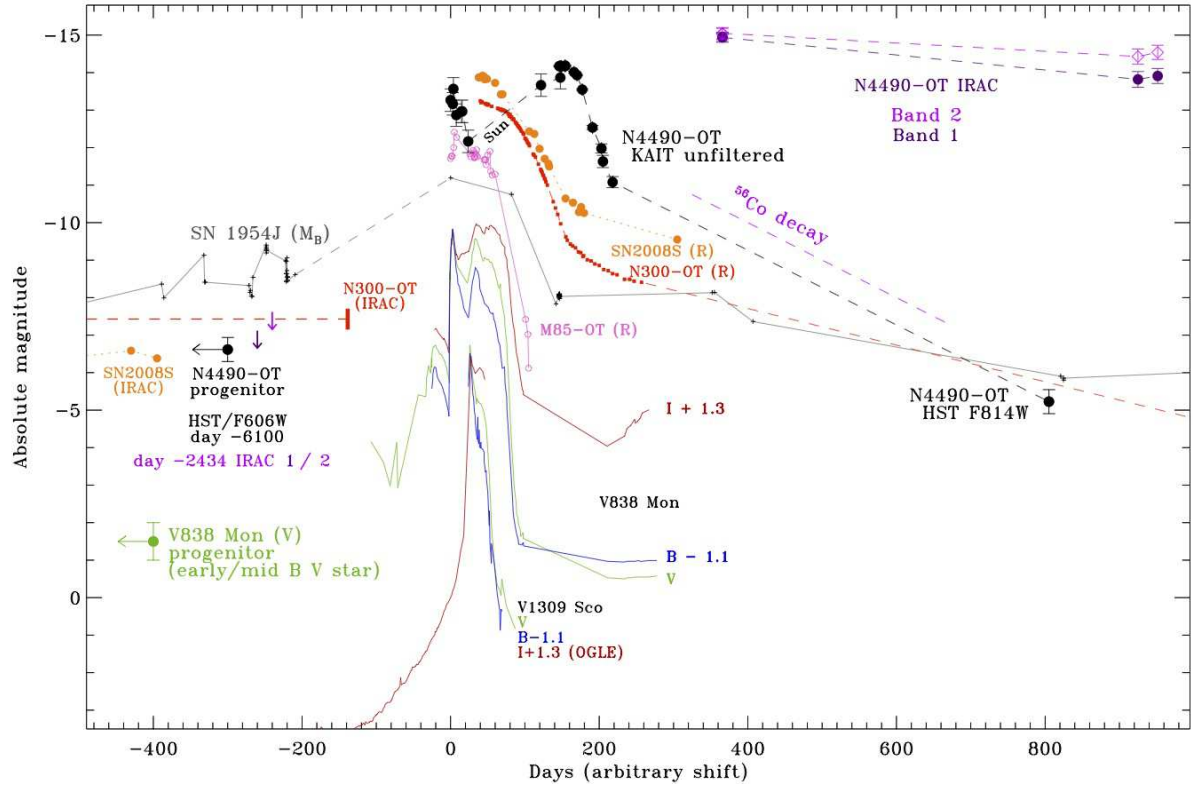}
\caption{Photometry of NGC~4490-OT and some related transients.  For
  NGC~4490-OT, we adopt the reddening and distance described in the
  text to place it on an absolute-magnitude scale for comparison with
  the other objects from the literature.  The black dots correspond to
  NGC~4490-OT, which includes unfiltered (approximately $R$-band)
  photometry taken from initial reports for the first few weeks plus
  KAIT unfiltered photometry after it reappeared from behind the Sun.
  The earliest point represents the progenitor that we measured in
  archival {\it HST} F606W images taken approximately 6100 days prior
  to discovery, as well as a late-time measurement with {\it HST} in
  the F814W filter.  See Table 1.  Photometry for the other objects is
  taken from the literature as follows: SN~2008S $R$ band
  \citep{smith+09} and IRAC \citep{prieto+08}, NGC~300-OT $R$ band
  \citep{bond09} and IRAC \citep{prieto08,thompson09}, SN~1954J $B$
  band \citep{ts68}, V838~Mon $BVI$ bands \citep{bond03,sparks08}, the
  likely V838 Mon progenitor assumed to be an early/mid B star for
  reasons noted in the text \citep{rr08,ab07}, and the V1309 Sco $I$
  band from OGLE \citep{tylenda11} and the $V$ and $B$ bands from
  AAVSO (the American Association of Variable Star Observers; see {\tt
    www.aavso.org.}).  The $R$-band light curve of M85-OT is shown in
  pink for comparison, from \citet{srk07}.}
\label{fig:phot}
\end{figure*}

\begin{table}\begin{center}\begin{minipage}{3.25in}
      \caption{KAIT Unfiltered Photometry of NGC~4490-OT}
\scriptsize
\begin{tabular}{@{}llccc}\hline\hline
Date  &MJD          &$m^a$       &$m-1\sigma$ &$m+1\sigma$ \\
...   &...          &(mag)     &(mag)       &(mag)   \\ \hline
2012 Jan 10 &55936.55     &15.793    &15.776      &15.810  \\
2012 Jan 12 &55938.40     &15.794    &15.778      &15.811  \\
2012 Jan 13 &55939.54     &15.799    &15.782      &15.816  \\
2012 Jan 18 &55944.48     &15.779    &15.763      &15.795  \\
2012 Jan 18 &55944.48     &15.795    &15.781      &15.809  \\
2012 Jan 30 &55956.42     &15.954    &15.933      &15.974  \\
2012 Feb 03 &55960.44     &16.035    &16.016      &16.053  \\
2012 Feb 10 &55967.47     &16.419    &16.399      &16.439  \\
2012 Feb 24 &55981.29     &17.430    &17.375      &17.486  \\
2012 Mar 07 &55993.35     &17.988    &17.868      &18.109  \\
2012 Mar 09 &55995.39     &18.336    &18.173      &18.500  \\
2012 Mar 22 &56008.34     &18.887    &18.739      &19.038  \\
2012 Apr 02 &56019.42     &(19.20)   &...         &...     \\
2012 Apr 07 &56024.42     &(18.56)   &...         &...     \\
2012 Apr 23 &56040.37     &(19.47)   &...         &...     \\
\hline
\end{tabular}\label{tab:kait}\end{minipage}
\end{center}
$^a$Upper limits are indicated in parentheses.
\end{table}

\section{OBSERVATIONS}

\subsection{Hubble Images}

A portion of NGC~4490 including the position of the transient was
observed by {\it HST} using the
Wide Field Planetary Camera 2 (WFPC2) on 1994 Dec. 3 as part of program
SNAP-5446 (PI G.\ Illingworth). The data consisted of a pair of
CR-split 80~s exposures in the F606W filter.  The location of the
transient fell on one of the larger WF chips with 0$\farcs$1
pixels. We obtained these pipeline drizzled images from the {\it HST}
archive.  This is the same image set that \citet{fraser11} used to
identify a probable source as the progenitor of NGC~4490-OT.  For this
source identified by them we measure $m$(F606W) = $23.58 \pm 0.24$
mag (1$\sigma$) using 0$\farcs$5 diameter aperture photometry
(Vega-based magnitudes).  This agrees within the uncertainty with the
magnitude reported by \citet{fraser11}.  With our adopted distance and
extinction, it corresponds to an absolute magnitude for the
progenitor of about $-$6.4 mag,
roughly 6100 days prior to discovery of the optical transient
outburst.  The {\it HST}/WFPC2 grey-scale image of the progenitor is shown 
in the upper-left panel of Figure~\ref{fig:hst}.

Here we confirm that this candidate was indeed the progenitor star of
the NGC~4490-OT transient, since later {\it HST} images show that the
same source has faded significantly after the outburst.  The same
position was imaged again with {\it HST} on 2013 Oct. 30, but this
time with the WFC3-UVIS camera and using several broadband filters
(see Table~\ref{tab:hstphot}).  These images were obtained as part of
the Legacy ExtraGalactic UV Survey (LEGUS) project (PI Calzetti;
GO-13364).  The filters and corresponding Vega-based magnitudes of the
same source are given in Table~\ref{tab:hstphot}.  Although these
images do not have the same F606W filter as the progenitor images,
both the F555W and F814W fluxes are fainter than the progenitor.  A
colour composite made from the F275W, F438W, and F814W images in 2013
Oct. is shown in the upper-right panels of Figure~\ref{fig:hst}.

We also investigate the host environment around the transient.
Photometry for sources in the environment of the OT was obtained from
the Level-0 source list generated by the LEGUS project (see
\citealt{calzetti15}). The individual LEGUS WFC3/UVIS `flt' frames
were first corrected for charge-transfer efficiency (CTE) losses using
the available online
tools\footnote{http://www.stsci.edu/hst/wfc3/tools/cte{\textunderscore}tools.}.
The corrected frames in each of the five LEGUS bands were then
processed with Dolphot \citep{dolphin00}, assuming a signal-to-noise
ratio threshold of 3 (and with CTE correction turned off). No further
selection criteria or cuts were imposed upon the data. The five-band
source lists were then bandmerged, with the OBJECT-TYPE flag set to
either `1' or `2' (i.e., a good star or one too faint for point-spread
function determination) and the ERRORFLAG $<1$. We used the Level-0
lists, rather than the Level-1 list, which have further cuts imposed
upon the photometry, in order to find as many sources in the OT
environment as possible. However, the Level-0 list almost certainly
contains a number of spurious sources. Following, for example,
\citet{gogarten09}, we considered star-like sources within a $\sim$100
pc radius of the OT, since stars formed in a common event will remain
spatially correlated up to this spatial scale for $\sim$100 Myr.
\citeauthor{gogarten09}~further limited the scale to $\sim$50 pc
around a source of interest as a compromise between including as many
coeval stars as possible, while limiting the amount of source
contamination.  We also consider this smaller spatial scale in our
analysis below.

Figure~\ref{fig:cc} compares colours from the photometry, particularly
in the ultraviolet (UV) and blue bands, to the colour-colour loci for
normal supergiant stars, and also assumes a reddening vector following
\citet{cardelli89} with $R_V=3.1$.  In Figure~\ref{fig:cc}, sources
within a 100 pc projected radius of NGC~4490-OT are green open
squares, and those projected within 50 pc are red open squares.
NGC~4490-OT itself is the filled black circle, although its UV colours
(especially including F275W) may be peculiar owing to line emission.
The implications for the local extinction toward NGC~4490-OT are
discussed below in Section 3.1.

\subsection{Optical Photometry}

The early-time light curve of NGC~4490-OT after discovery is poorly
sampled, because the object became difficult to observe owing to hour-angle
constraints a few weeks after discovery.  In Figure~\ref{fig:phot} we
collect early unfiltered CCD magnitudes reported by
amateur observers.\footnote{See
  http://www.rochesterastronomy.org/supernova.html} Since these are
not all on the same photometric system, we adopt representative error
bars of $\pm0.4$ mag for the purpose of characterizing the early
photometric behaviour.  We emphasize, however, that these are not our
measurements and we do not have adequate information to assess their
true uncertainty; we only show them for comparison. Measurements by several
different amateur observers agree that the source faded by $\sim1.5$
mag in the first few weeks.

Our photometric monitoring of NGC~4490-OT commenced when the object
returned from behind the Sun in January 2012. We used the Katzman
Automatic Imaging Telescope (KAIT; \citealt{filippenko01};
\citealt{filippenko03}) at Lick Observatory to obtain unfiltered
photometry. As demonstrated by \citet{li02}, the best match to
broadband filters for the KAIT unfiltered data is the $R$ band.  We
list the KAIT apparent unfiltered (approximately $R$) magnitudes of
NGC~4490-OT in Table~\ref{tab:kait}. To put the flux on an
absolute-magnitude scale, we adopt the distance and reddening listed
in the Introduction; the resulting absolute-magnitude light curve is
shown in Figure~\ref{fig:phot}, along with other transient sources
from the literature for comparison.

\begin{table}\begin{center}\begin{minipage}{3.25in}
      \caption{Spitzer Photometry of NGC~4490-OT}
\scriptsize
\begin{tabular}{@{}llcccc}\hline\hline
Date &MJD          &IRAC1     &1$\sigma$ &IRAC2    &1$\sigma$ \\
... &...          &(mag)     &(mag)     &(mag)    &(mag)   \\ \hline
%
2004 Dec 17 &53356.04     &(22.8)   &...       &(22.3)  &...     \\  
2012 Aug 17 &56156.03     &14.97     &0.12      &14.89    &0.15     \\ 
2012 Aug 17 &56156.18     &14.98     &0.13      &14.87    &0.15     \\ 
2014 Feb 27 &56715.32     &16.10     &0.21      &15.49    &0.20     \\ 
2014 Mar 25 &56741.84     &16.01     &0.20      &15.38    &0.19     \\ 
\hline
\end{tabular}\label{tab:spirits}\end{minipage}
\end{center}
Upper limits are indicated in parentheses.
\end{table}

\subsection{IR Imaging}

We observed the host galaxy NGC~4490 in the IR as part of
the large program called SPitzer InfraRed Intensive Transients Survey
(SPIRITS; PI M. Kasliwal), conducted using the Infrared Array
Camera (IRAC; \citealt{fazio}) Bands 1 and 2 (3.6 and 4.5 $\mu$m,
respectively) during the {\it Spitzer Space Telescope} ({\it Spitzer})
warm mission in Cycle 10.  Our observations were obtained on 2014 Feb.
27 and Mar. 25.  A source coincident with the NGC~4490 optical
transient was identified in the IR by comparison with archival {\it
  Spitzer} images obtained in the same filters in 2004 and 2012.  No
source was detected in the first epoch in 2004, so we used this set of
images as a template to subtract from the 2012 and 2014 images.  The
resulting Band 1 and 2 magnitudes for various dates measured on
template-subtracted images are given in Table~\ref{tab:spirits}.  For
the nondetection on the first epoch in 2004, we list upper limits in
Table~\ref{tab:spirits}.  These are 5$\sigma$ upper limits based on
the background noise measured in an annulus around the position of the
source in the template images and the detection images added in
quadrature.  Experimenting with annuli that had a variety of inner and
outer radii gave consistent results within $\pm0.1$ mag.

These IR magnitudes from detections in template-subtracted images, as
well as 2004 progenitor upper limits, are plotted in
Figure~\ref{fig:phot}. The date of the progenitor upper limits from
{\it Spitzer} corresponds to day $-$2434, which is far off the left
side of the plot in Figure~\ref{fig:phot}, so the limits are shown
shortly before the transient's discovery for reference.  For the IR
absolute magnitudes, we assumed the same distance modulus of
$m-M=29.92$ mag as adopted above for optical data.  The upper limits
to the progenitor's IR luminosity are comparable to the luminosity of
the progenitor source detected in {\it HST} images.  Thus, although we
did not detect the IR flux of the progenitor itself, we can see that
it is clearly not a heavily obscured progenitor dominated by IR
emission as in the cases of SN~2008S and the 2008 transient in
NGC~300.  The optical-to-IR spectral energy distributions (SEDs) of
the progenitor and late-time source are discussed later.



We also obtained a $K_s$-band image of the site of NGC~4490-OT on 2014
May 10, using WIRC on the Palomar 5-m telescope.  This corresponds to
day 997 after discovery, and is similar in time to our last epoch of
{\it Spitzer} Band 1 and 2 photometry.  The image was calibrated using
2MASS stars in the field, and we derive a $K$-band brightness of
$18.7\pm0.2$ mag.

\begin{table}\begin{center}\begin{minipage}{3.25in}
      \caption{Spectroscopic Observations of NGC~4490-OT}
\scriptsize
\begin{tabular}{@{}llcccc}\hline\hline
Date     &Tel./Instr &Day$^a$ &$\Delta\lambda$(\AA) &note &$T_{\rm BB}$(K) \\ \hline 
2012\,Jan.\,18 &Lick/Kast     &154   &3500--10,000  &...     &5100   \\  %
2012\,Jan.\,19 &MMT/BC        &155   &5700--7000   &...     &...   \\   %
2012\,Jan.\,29 &MMT/BC        &165   &3800--9000   &clouds  &...   \\   %
2012\,Feb.\,01 &Lick/Kast     &168   &3500--10,000  &...     &4700   \\  %
2012\,Mar.\,01 &MMT/BC        &197   &5700--7000   &clouds  &...   \\   %
\hline
\end{tabular}\label{tab:spectab}\end{minipage}
\end{center}
$^a$Day refers to approximate days after discovery, 2011 Aug 16.83.
\end{table}

\begin{figure*}
\includegraphics[width=5.8in]{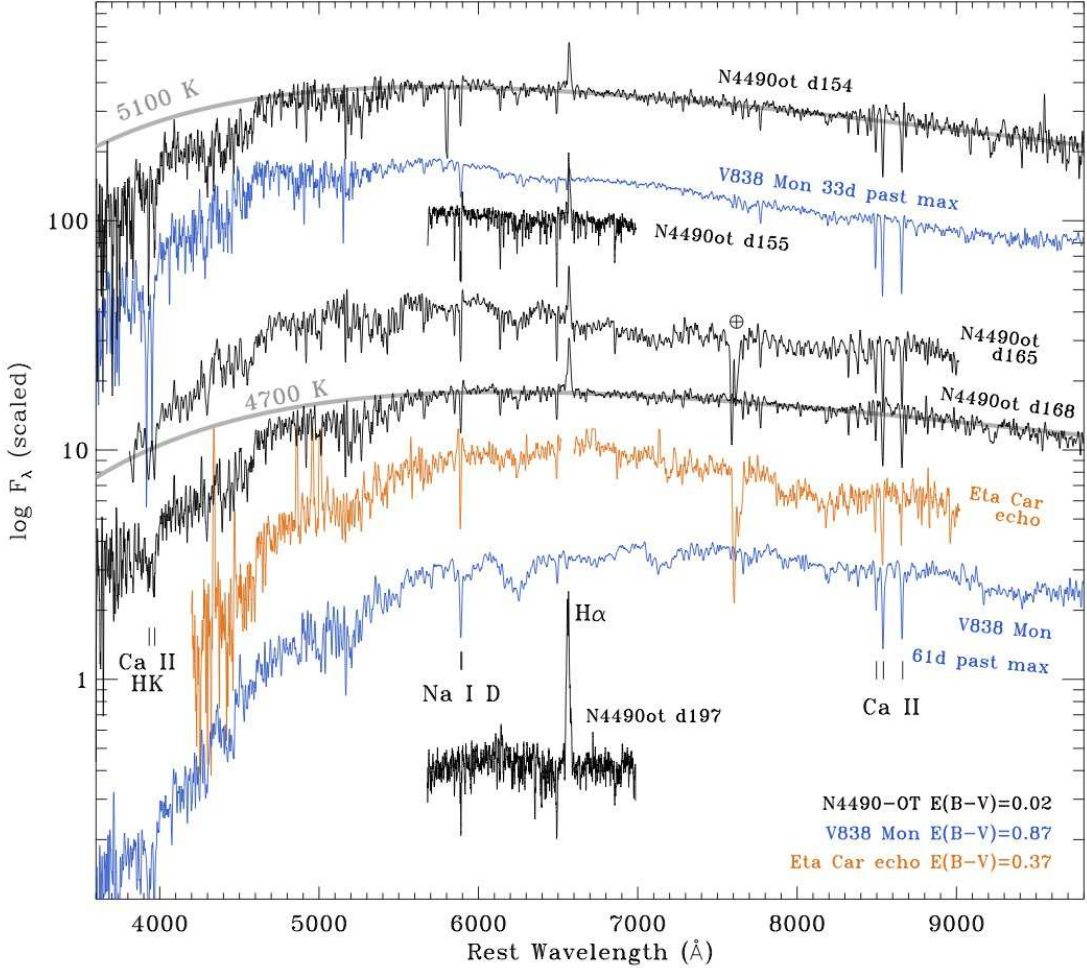}
\caption{Optical spectra of N4490-OT compared to some spectra of V838
  Mon.  Details of our N4490-OT spectra are given in
  Table~\ref{tab:spectab}, and in this plot the spectra have been
  dereddened by a line-of-sight reddening value of $E(B-V)=0.019$ mag.
  The spectra of V838~Mon were obtained with the Kast spectrograph at
  Lick Observatory on 2002 Mar. 11 and Apr. 8 (days 33 and 61,
  respectively, past the $V$-band peak of V838 Mon; see
  \citealt{munari02}).  These V838~Mon spectra were taken from the
  Berkeley SN spectral database and have been discussed previously by
  \citet{smith11}.  The V838~Mon spectra are dereddened by
  $E(B-V)=0.87$ mag \citep{munari02,munari07}. Although the days past
  maximum brightness of the V838~Mon spectra are not chronologically
  in step with the N4490-OT spectra, we chose to stagger them this way
  because N4490-OT had a much longer duration, and so these spectra
  represent comparable epochs during the decline from peak.  We also
  include a spectrum of a light echo from $\eta$ Carinae's giant
  eruption \citep{rest12}, dereddened by $E(B-V)=0.37$ mag as
  appropriate for our line of sight to the Carina Nebula
  \citep{smith04b}.}
\label{fig:spectra}
\end{figure*}

\subsection{Optical Spectroscopy}

We obtained five epochs of visible-wavelength spectroscopy during the
sharp decline after the second luminosity spike that occurred around
day 150; this also coincided with the time when NGC~4490 returned from
behind the Sun and when our KAIT photometric monitoring began.  We
used the Bluechannel (BC) spectrograph on the 6.5-m Multiple Mirror
Telescope (MMT) and the Kast spectrograph \citep{1993ms} on the Lick
3-m Shane reflector, with observation details given in
Table~\ref{tab:spectab}.  The slit was always oriented at the
parallactic angle \citep{filippenko82}, and the long-slit spectra were
reduced using standard procedures.  Final spectra are shown in
Figure~\ref{fig:spectra}.  Some of the comparison spectra in
Figure~\ref{fig:spectra} were taken from the Berkeley SN spectral
database\footnote{http://heracles.astro.berkeley.edu/sndb/} as noted
in the caption.  The Lick/Kast spectra covered a wide wavelength range
at low resolution in a blue and a red channel, while the MMT/BC
spectra either covered a wide range in a single order (300 line
mm$^{-1}$ grating) or a smaller wavelength range with higher
resolution (1200 line mm$^{-1}$ grating) including Na~{\sc i}~D
$\lambda\lambda$5890, 5896 and H$\alpha$.  For days 154 and 168 in
Figure~\ref{fig:spectra} we show a representative blackbody for
comparison.  All epochs exhibit strong line blanketing in the blue,
and the characteristic temperature gets cooler with time as the
transient fades rapidly.  Narrow H$\alpha$ emission and the Ca~{\sc
  ii} near-IR triplet in absorption remain prominent at all epochs.

In our moderate-resolution MMT spectrum taken on day 155 near the time
of peak luminosity, we measure a total Na~{\sc i}~D equivalent width
(EW) of $4.3\pm0.6$~\AA.  This is the total EW, including Galactic
absorption and absorption in the local host galaxy (at the redshift of
NGC~4490), as well as absorption in the photosphere of the transient;
these components are blended at our observed resolution.  Examining
the spectra, the strong Na~{\sc i} absorption is blueshifted
absorption that forms part of a P-Cygni profile with narrow Na~{\sc
  i}~D emission.  It is therefore likely that a large fraction of this
absorption occurs in the wind/atmosphere of the transient source, and
is therefore not a reliable measure of any local extinction.  The
Milky Way reddening value is $E(B-V)=0.019$ mag, as noted above, but
there may be additional local host-galaxy reddening.  A value of
$E(B-V)=0.3$ mag for additional host reddening was inferred for the
SN~2008ax by \citet{pastorello08}, which is located nearby in the same
galaxy, and below we discuss the likely value for the specific
location of NGC4490-OT based on photometry of stars in the local
environment.  Any increase in the assumed reddening would increase our
estimates of the peak luminosity of the transient, the inferred
luminosity and mass of the progenitor, and the brightness of the
dust-obscured source at visible wavelengths, but it would not change
our estimate of the IR luminosity seen at late times.

\subsection{Near-IR Spectroscopy}

On 2014 June 7 we obtained a near-IR $K$-band spectrum at the position
of NGC~4490-OT in its late-time aftermath of the outburst, using
MOSFIRE on Keck.  Although {\it Spitzer} data indicate a strong
thermal-IR excess at this epoch, we did not detect continuum emission
or any emission lines in our MOSFIRE spectrum with an exposure time of
1074\,s.

\section{DISCUSSION}

\subsection{Extinction and Environment}

As is often the case, the nature of the progenitor star that one
infers from a single-filter {\it HST} detection is clouded by the
possibility of large and uncertain local extinction in the host galaxy.
Local dust --- especially circumstellar dust that surrounded the
progenitor --- may have been vaporised by the transient event, and
there may be additional ISM extinction in this particular region of
the host galaxy.  To assist in our analysis of the progenitor below,
we consider stars in the surrounding environment of NGC~4490-OT for
additional constraints on the possible and likely local ISM
extinction.

Figure~\ref{fig:cc} compares colours from the {\it HST} LEGUS
photometry, particularly in the UV and blue bands, to the
colour-colour loci for normal supergiant stars, and also assumes a
reddening vector following \citet{cardelli89} with $R_V=3.1$.  In
Figure~\ref{fig:cc}, sources within a 100 pc projected radius of
NGC~4490-OT are green open squares, and those projected within 50 pc
are redopen squares.  For reference, the blue dashed line shows a
reddening vector corresponding to $A_V=3$ mag.

The resulting plot shows a great deal of scatter, indicating patchy
and sometimes large extinction in the surroundings.  Some stars in the
surrounding 100 pc (green and red) have colours consistent with little
or no reddening, while some show colours that suggest as much as 2 mag
of $V$-band extinction.  Indeed, it is apparent from the colour image
in Figure~\ref{fig:hst} that NGC~4490 is speckled with patchy
extinction.  This holds true when we restrict ourselves only to the
few sources within 50 pc (red).  In the left panel of
Figure~\ref{fig:cc} (F336W--F438W vs.\ F438W--F555W), one of the two
red points at the top right of the plot corresponds to a star only a
few (6--8) pixels from NGC~4490-OT, and this star has a colour
consistent with $A_V \approx 2$ mag.  This might suggest rather large
extinction in the immediate environment of the transient.  On the
other hand, many of the other stars have a range of values from $A_V
\approx 0$ to $A_V \approx 2$ mag. Since choosing between these
remains uncertain, in our analysis of the progenitor below we consider
the impact of three different options for the adopted local reddening
in addition to the Milky Way line-of-sight reddening.  Option (1)
corresponds to no additional local reddening as a lower limit, for
option (2) we consider $A_V=1.0$ mag as a likely value, and for option
(3) we consider $A_V=2.0$ mag as a likely upper limit to the local
reddening.

\subsection{Progenitor and Aftermath}

The late-time {\it HST} images show a source at the same position as
the progenitor, but fainter.  This fading confirms the candidate
progenitor source first identified by \citet{fraser11}.\footnote{Of
  course, some of the light contributing to this source could also be
  from a companion star, which may have been engulfed by the expanding
  dusty ejecta as in the case of V838~Mon, but in any case it is
  likely to represent the integrated flux of the system that gave rise
  to the transient source.}

\subsubsection{Progenitor}

We measure an F606W absolute magnitude of $-6.4 \pm 0.24$ for the
progenitor around 6100 days before the transient was discovered
(corrected only for Milky Way reddening).  The pre-eruption {\it HST}
data include only a single filter and only one epoch, but this offers
some useful constraints on the possible nature of the progenitor.
With no attempt at a bolometric correction, the progenitor system must
be at least as luminous as the progenitor of SN~2008S (and almost as
luminous as the progenitor of NGC~300-OT).  Unlike SN~2008S and
NGC~300-OT, however, this detection is at visible wavelengths, so the
progenitor is not as heavily obscured by its own circumstellar dust.

On the other hand, the progenitor could also be blue, like the
progenitor of SN~1954J and other LBVs.  In this case the F606W
absolute magnitude of $-$6.4 mag is a lower limit to the bolometric
luminosity, because a bolometric correction or extinction correction
must raise the progenitor's bolometric luminosity.

In fact, we favour the interpretation that the progenitor was a more
luminous blue star, based largely on the IR nondetection of the
progenitor in {\it Spitzer} data.  The IR nondetections actually
place strong constraints, provided that the star did not change
substantially between 1994 ({\it HST} F606W image) and 2004 ({\it
  Spitzer} Bands 1 and 2).  Of course, the star might have changed its
luminosity and colour somewhat in this time if it was out of
equilibrium and approaching a merger or some other disruptive event,
so we can only take this information as approximate (moreover, if
V1309~Sco or LBV eruptions like $\eta$ Car provide guidance, the star
may have been brightening and getting cooler as it approached its
transient event, which would reinforce the analysis below).  With this
potential caveat aside, the {\it Spitzer} nondetections of the
progenitor are quite constraining.

\begin{figure}
\includegraphics[width=2.9in]{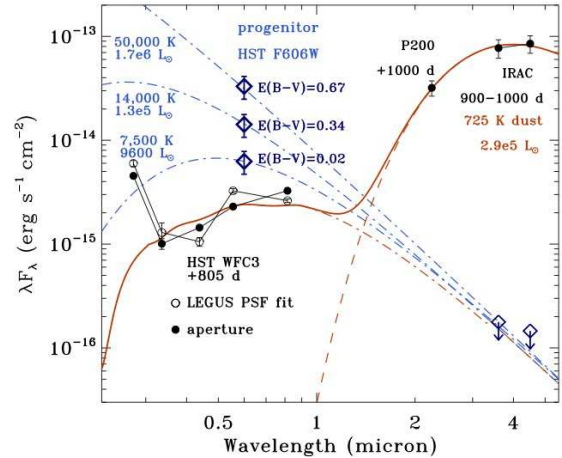}
\caption{Observed SED of the progenitor of NGC~4490-OT (blue diamond
  symbols, including different reddening corrections; see text), as
  well as the source at the same position at late times (black circles
  and orange models). For the {\it HST} measurements of the late-time
  source, we show the results of PSF fitting photometry from the LEGUS
  photometric catalog (unfilled black circles) as well as aperture
  photometry using a 0$\farcs$5-diameter aperture (filled circles).}
\label{fig:sed}
\end{figure}

Figure~\ref{fig:sed} shows the SED of the candidate progenitor of
NGC~4490-OT, as well as its aftermath (described below).  We plot the
{\it HST} F606W detection in 1994, plus the upper limits to $\lambda
F_{\lambda}$ from our {\it Spitzer} IRAC Band 1 and 2 nondetections
in 2004 (open blue diamonds).  For the {\it HST} F606W detection of
the progenitor we must assign an extinction correction, but this is
uncertain, as noted above.  Following an analysis of stars in the
local environment of the transient, we consider three options for the
extinction correction, and we plot all three of these on the SED in
Figure~\ref{fig:sed}.  They are discussed below, along with
implications for the nature of the progenitor.

{\it Option 1:} Here we correct only for the adopted line-of-sight
Milky Way reddening of $E(B-V)=0.02$ mag with zero additional
reddening in the host galaxy.  This is regarded as a lower limit.  A
blackbody through the F606W point and below the IRAC limits would need
to have a temperature of at least 7500~K and a luminosity of more than
9600\,L$_{\odot}$.  This would correspond to a relatively
low-luminosity F-type supergiant or giant from an intermediate-mass
star.  This luminosity is more than an order of magnitude lower than
the post-outburst IR luminosity.

{\it Option 2:} Next we consider adding $A_V=1.0$ mag of local
extinction ($E(B-V) = 0.32$ mag), so the total reddening is $E(B-V) =
0.34$ mag.  We regard this reddening correction as a likely value, and
it is similar to the value adopted for SN~2008ax by
\citet{pastorello08}.  This F606W point and the IRAC limits would
correspond to a blackbody with a temperature of at least 14,000~K
(late B-type) and a luminosity of $1.28\times10^5$ L$_{\odot}$ or
more.  Such a star might be very similar to the class of
lower-luminosity LBVs in their quiescent state, like HR Carinae or
HD~168625 \citep{smith04,smith11}. They have effective initial masses
(compared to single-star models) of $\sim30$ M$_{\odot}$.  For LBVs,
there are of course significant caveats to an ``initial mass'' derived
from single-star evolution models in cases where binary mass transfer
or mergers may be relevant (see \citealt{st15}).  Since the reddening
is very uncertain, the temperature and luminosity could easily be
somewhat more or less, but more reddening would simultaneously require
the star to be more luminous and hotter, moving it roughly along the
S~Dor instability strip.  This makes it seem plausible that the
progenitor star was indeed a quiescent LBV.  Interestingly, the
luminosity of such a progenitor is within a factor of 2--3 of the
post-outburst IR source's luminosity, which would be consistent with
the star surviving the event or with a merger product that is even
more luminous than the progenitor.  Interpreting the progenitor as a
lower-luminosity LBV would also seem to be consistent with the
relatively sparse environment in the vicinity of NGC~4490-OT (i.e.,
not in a massive cluster or H~{\sc ii} region), since these
lower-luminosity LBVs tend to be tens to hundreds of pc from clusters
of O-type stars \citep{st15}.

{\it Option 3:} Next we consider adding $A_V=2.0$ mag of local
extinction ($E(B-V) = 0.65$ mag), based on the extinction inferred for
the detected star nearest to the transient source, so the total
reddening is $E(B-V) = 0.67$ mag.  We regard this as a probable upper
limit for the local ISM reddening within the host galaxy, although
there may also have been considerable circumstellar extinction that
was destroyed by the transient event.  A blackbody through this F606W
point and below the IRAC limits would need to be a very hot and
luminous star, with $T \ga 50,000$~K and $L \ga 1.7 \times 10^6$\,
L$_{\odot}$.  Such a star would be a very luminous early O-type
main-sequence star or a WNH star with an initial mass around 80--100
M$_{\odot}$.  Given the relative isolation of NGC~4490-OT in its host
galaxy, this seems less likely than Option 2.

In any of the three options above, the IR upper limits do not allow
the source detected by {\it HST} to be a cool evolved star like an
asymptotic giant branch star or red supergiant.  We therefore consider
it quite plausible that the progenitor was a traditional LBV or
related class of evolved luminous blue star (Option 2).  Options 1 and
3 seem much less likely for the reasons noted above.  In any case, it
is interesting that the progenitor of N4490-OT is substantially more
luminous and more massive than the expected progenitor systems of V838
Mon (roughly $-$1.5 mag; B-type star) or the low-mass progenitor of
V1309 Sco (roughly $+$4.8 mag).

\subsubsection{Aftermath}

The position of the transient was observed again at late times with
{\it HST}/WFC3, 805 days after discovery, and this time we have more
colour information from multiple {\it HST} filter detections from the
UV to near-IR.  We also have a ground-based near-IR detection in the
$K$ band, plus clear detections with IRAC Bands 1 and 2 at four epochs
between 300 and 1000 days after discovery.  These late-time
measurements of the SED are also plotted in Figure~\ref{fig:sed},
where we choose the IRAC Band 1 and 2 photometry closest in time to
the {\it HST} photometry; we do not see evidence of strong IR
variability at 900--1000 days, only a gradual fading. For the
late-time {\it HST} photometry, we show two measurements of the same
data.  The unfilled circles correspond to PSF-fitting photometry from
the LEGUS photometric catalog (see Table~\ref{tab:hstphot}, while the
filled circles represent magnitudes measured using a
0$\farcs$5-diameter aperture.

In Figure~\ref{fig:sed}, the observed photometry of the late-time
source (black filled circles) is compared to a simple two-component model
with a reddened hot star and warm dust.  The reddened star is the same
14,000~K blackbody matched to the progenitor in Option 2 above, but
reddened by $E(B-V)=0.7$ mag.\footnote{Of course, one would not
  necessarily expect the post-outburst star to have the same
  temperature as the progenitor, so this is merely illustrative.}  The
dust component is a single-temperature 725~K grey body ($\lambda^{-1}$
emissivity) that fits the $K$-band and IRAC points reasonably well.
Given that this two-component model is probably quite oversimplified,
the resulting combined model gives a fair representation of the
observed photometry excluding the F275W point.  The F275W measurement
is admittedly somewhat strange; the F275W--F336W colour is steeper than
the Rayleigh-Jeans tail of an unreddened blackbody, so it seems
likely that the F275W filter has some strong contamination from
Mg~{\sc ii} $\lambda$2800 line emission.  Except for this point, it
appears plausible that the post-outburst source is a luminous hot star
enshrouded by its own circumstellar dust that was formed in the
eruption event.  In this sense, it conforms to traditional
expectations for a giant LBV eruption.

\subsubsection{Late-time IR (non)variability}

The fact that the IRAC Band 1 and 2 fluxes have not dropped more
substantially at late times is physically significant.  A strong IR
excess at late times indicates either pre-existing circumstellar medium
(CSM) dust or dust
that was newly formed in material ejected during the outburst.  In any
case, the dust could be heated by a luminous post-outburst source
(indicating that the star likely survived the event), or it could be
an ``IR echo'' wherein surrounding dust is heated by the pulse of
UV/optical radiation from the transient event itself, and then
re-emits its absorbed UV/optical energy in the thermal IR.

For an IR echo, the maximum dust temperature and the dust luminosity
should fall with time as the light echo expands (see, e.g.,
\citealt{fox11,fox15}).  The hottest dust will be found at the point
of the light-echo paraboloid that is closest to the source, which is
on the far side of the transient source in a spherical dust shell.
This is at a radius of $r = ct/2$, where $c$ is the speed of light and
$t$ is the time since the UV/optical peak of the transient source.
For a pulse with peak luminosity $L$, the grain temperature of the
hottest echo-heated dust at $r$ is then given by

\begin{displaymath}
  T_d^4 = (\frac{Q_{\rm UV}}{Q_{\rm IR}}) \frac{L}{16 \pi r^2 \sigma},
\end{displaymath}

\noindent 
where $Q_{\rm UV}$/$Q_{\rm IR}$ describes the grain efficiency in the
UV (absorption) and IR (emission), and $\sigma$ is the
Stefan-Boltzmann constant.  At $t \approx 900$--1000~d after the peak,
when our $K$-band and second epoch of IRAC data were obtained, the
minimum radius of the echo was $r \approx 0.8$ pc.  At that radius the
observed dust temperature of $\sim725$~K would require a transient
peak luminosity of roughly $1.1 \times 10^{12}$\, L$_{\odot}$ $\times
(Q_{\rm UV}/Q_{\rm IR})$.  We dont know the grain properties {\it a
  priori}, but assuming a typical value of $Q_{\rm UV}/Q_{\rm IR}
\approx 0.3$ would require an absolute magnitude of $-$23.6 mag.  This
is more luminous than any known SN, and of course vastly more luminous
than the observed IR source, so we conclude that the observed colour
temperature of the IR excess at late times is too hot to be explained
by an IR echo.  The dust must be located much closer to a central
engine.  Moreover, the temperature of the hottest echo-heated dust
will fall as $T \propto t^{-0.5}$.  At the time of our first epoch of
IRAC observations around day 360, the maximum dust temperature should
have been around 1200~K.  This would shift the peak of the SED from 4
$\mu$m at day 1000 to around 2.4 $\mu$m at the earlier epoch.
However, there is little change in the colour and IR luminosity
between our IRAC epochs, adding further doubt to the IR-echo
hypothesis.

On the other hand, the roughly constant post-outburst IR luminosity is
close to the reddening-corrected progenitor star's luminosity for our
favoured Option 2 above.  The factor of $\sim2$ luminosity difference
between the progenitor model of a 14,000 K star and the late-time IR
luminosity could easily be accounted for by (1) a slightly higher
temperature and luminosity for the progenitor for the same reddening
(the IRAC upper limits only provide a lower limit to the star's
temperature) or a somewhat higher reddening, (2) some contribution
from an IR echo or residual luminosity from the fading eruption, or
(3) an actual increase in the luminosity of the post-outburst source
as one might expect if it were a stellar merger or if there was some
other energy injection into the star's envelope.  The pre- and
post-outburst luminosities are, however, of the same order of
magnitude, making it plausible that the relatively constant late-time
IR excess is caused by a dust shell heated by a central surviving
star.

For the IR luminosity of the dust shell, $L = 2.9 \times 10^{5}\, {\rm
  L}_{\odot}$, an equilibrium grain temperature of 725~K in a
spherical shell corresponds to a radius of roughly 80 AU.  To reach
this radius by a time of $\sim900$ d after ejection would require an
expansion speed of roughly 160 km s$^{-1}$.  Of course, small grains
may be hotter than the equilibrium blackbody temperature, so the
implied radius and required ejection speed may be a factor of 2-3
higher.  As noted below, we infer ejection speeds from spectra of a
few hundred km s$^{-1}$.  Given factors of 2 or more owing to
geometric uncertainty (we see evidence for asymmetry, as noted below),
the observed IR SED at late times is reasonably consistent with
expectations for emitting warm dust that formed from mass ejected in
the 2011--2012 transient event.  It is less likely that this is
pre-existing dust, since dust at this same radius ($\sim$80 AU) would
have been heated to a temperature of around 2200~K by the peak
luminosity of the transient, and would therefore have been largely
destroyed.

For standard assumptions about the grain density, an IR luminosity of
$2.9 \times 10^5$ L$_{\odot}$ for 725~K dust requires an emitting mass
of, very roughly, $10^{-9}$\, M$_{\odot}$ (see, e.g.,
\citealt{smith03}).  The uncertainty may be a factor of 2 or more
depending on assumptions about the grains, but then again, this is
merely a lower limit to the dust mass since there may be a much larger
mass of cool dust present and emitting primarily at far-IR wavelengths
and unconstrained by our 3--5 $\mu$m data.  The near-IR excess
therefore gives little useful information about the total mass ejected
in the event.

\subsection{Visible-Wavelength Light Curve}

From initial unfiltered magnitude estimates reported by amateur astronomers,
NGC~4490-OT showed an initial peak at an absolute magnitude of roughly
$-$13.5, followed by a fast decline in the weeks after discovery
(Figure~\ref{fig:phot}).  It then became unobservable because it was
too close to the Sun during days 30--120 after discovery.  When we
began our unfiltered KAIT photometric monitoring after the source
became observable again, we found it to be brighter than at discovery,
and was rising again to a second peak at $\sim150$--160 d, reaching
an absolute unfiltered (approximately $R$-band) peak of $-$14.2 mag
(correcting only for Milky Way extinction).

The source then faded rapidly for $\sim60$~d after this second
peak.  Subsequently, the decline slowed, and the average
(interpolated) decline rate between days 220 and 800 was similar to
the $^{56}$Co decay rate (Figure~\ref{fig:phot}), although we do not
have data to characterize the details of the decline in this time
interval.  Unfortunately, we did not obtain multi-filter photometry to
document the optical colour evolution during and after peak.  However,
our visible-wavelength spectra (see below) do show increasingly red
colours and an evolution to cooler apparent temperatures during this
time, as the source faded from peak.  In fact, the colours became quite
red at the latest epochs observed by {\it HST} at day 805, with a
strong IR excess.

The light curve of NGC~4490-OT, with multiple peaks and an evolution
to red colours and IR excess, is qualitatively reminiscent of the
complex multi-peaked light curves observed in some Galactic transients
that have been described as stellar merger events.  Both V838~Mon and
V1309~Sco show an initial blue peak with a rapid decline on a time
scale similar to that of NGC~4490-OT, and a comparably luminous second
or third peak after that with increasingly red colour.  The light
curves of these two objects are shown in Figure~\ref{fig:phot}.  An
important difference in the light curves, however, is that NGC~4490-OT
had a much higher peak luminosity, and a longer duration.  V1309 Sco
faded in about one month and V838 Mon faded in 80--90 days, while
NGC~4490-OT stayed bright for $\sim$200 days.  Interestingly, the
duration and brightness of these events appears to correlate with the
suspected mass of their progenitor stars, as we discuss later.  The
19th century Great Eruption of $\eta$ Carinae also showed multiple
narrow peaks followed by a longer broad plateau \citep{sf11}, and its
duration was a decade or more. Since $\eta$ Car is likely to be a much
more massive star, it may extend this same correlation.

The bright phase of the transient (days 0--200) is poorly sampled
because the source's position in the sky was near the Sun for much of
this time.  Simply interpolating between observation dates and
integrating the optical luminosity can give a crude estimate of the
total radiated energy (ignoring any bolometric correction) of very
roughly $E_{\rm rad} \approx 1.5 \times 10^{48}$ ergs.  This is less
than $\eta$ Car's giant eruption, but substantially more than other
less-energetic LBV eruptions like P Cygni \citep{sh06,smith11} or
SN~1954J \citep{hds99}.  It is also orders of magnitude more than the
lower-mass stellar merger sources mentioned above.

\begin{figure}
\includegraphics[width=2.9in]{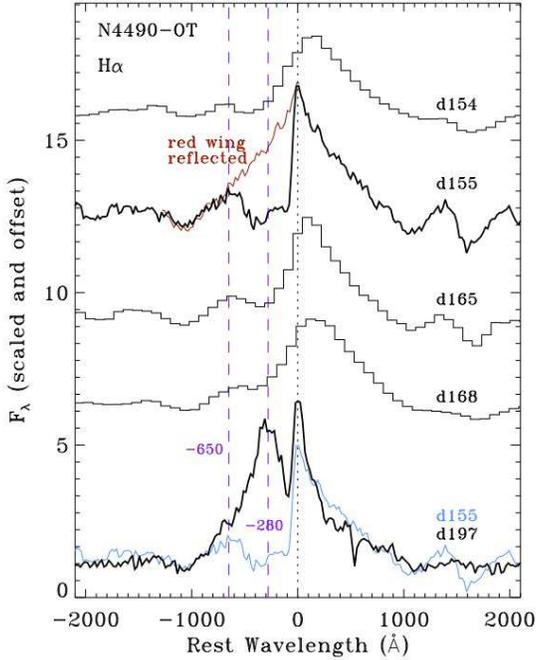}
\caption{Details of the H$\alpha$ line profiles in our spectra of
  NGC~4490-OT.  The two MMT spectra on days 155 and 197 have higher
  spectral resolution than the other three.  The spectrum in red
  plotted over the day 155 spectrum shows what the H$\alpha$ line
  profile would look like if it were symmetric, by taking the red wing
  of the line and reflecting it to the blue side of the line.  Also,
  the day 155 spectrum is reproduced in blue and plotted over the last
  epoch on day 197 for comparison, showing that the blueshifted
  absorption on day 155 occurred at similar velocities to the excess
  blue emission on day 197.  Representative velocities of $-$650 km
  s$^{-1}$ (blue edge of the absorption) and $-$280 km s$^{-1}$
  (middle of the absorption trough and emission peak) are shown with
  purple dashed lines.}
\label{fig:halpha}
\end{figure}

\subsection{Spectroscopic Evolution}

Figure~\ref{fig:spectra} shows the evolution of the visible-wavelength
spectrum of NGC~4490-OT.  Unfortunately, we were not able to obtain any
early-time spectra.  As noted above, however, \citet{magill11}
reported that early-time spectra showed a blue continuum with bright
narrow Balmer emission lines reminiscent of LBVs.  Our spectra
document times just before the second peak, which is the time of
maximum observed luminosity, and the rapid decline from this main
peak.  At these phases, NGC~4490-OT showed a warm continuum consistent
with temperatures around 5000~K, accompanied by weak H$\alpha$
emission, strong Ca~{\sc ii} absorption (both the H\&K lines and the near-IR
triplet), and strong metal line-blanketing absorption in the blue.

This apparent temperature around 5000~K is similar to that indicated
by spectroscopy of light echoes from $\eta$ Carinae's 19th century
eruption \citep{rest12}.  Similar absorption features were also seen
in the light echo spectra of $\eta$ Car and some extragalactic LBVs
like UGC~2773-OT \citep{rest12,smith11,smith16}.

Interestingly, the spectra of NGC~4490-OT do not show the forbidden
[Ca~{\sc ii}] $\lambda\lambda$7291, 7323 doublet in emission (or the
Ca~{\sc ii} near-IR triplet in emission), which is another way that
NGC~4490-OT seems distinct from SN~2008S-like objects
\citep{prieto+08,thompson09,bond09,berger09,smith+09}.  The spectra of
NGC~4490-OT during its decline do, however, bear a striking
resemblance to spectra of the Galactic transient V838 Mon during its
decline from peak luminosity \citep{munari02,munari07}.
Figure~\ref{fig:spectra} compares our spectra of NGC~4490-OT to those
of V838~Mon.  (While the times after peak of 30--70 days for V838~Mon
are different than for the day 150--200 spectra of NGC~4490-OT, these
trace similar epochs during the decline because NGC~4490-OT had a much
longer timescale, fading at 180--200 days instead of $\sim 80$ days
for V838~Mon). Spectra of both objects show a similar evolution in
apparent continuum slope, they both exhibit a similar forest of narrow
atomic absorption lines and line blanketing in the blue that are
characteristic of yellow F and G-type supergiants, and both have very
strong Ca~{\sc ii} absorption in the H\&K lines, as well as strong
absorption in the Ca~{\sc ii} near-IR triplet.

A notable difference is that V838 Mon did not exhibit strong H$\alpha$
in emission during these epochs, whereas NGC~4490-OT did. (V838~Mon
did, however, show H$\alpha$ emission at some earlier epochs;
\citealt{munari07}.)  The emission EW rises from about $-14$ \AA \ at
peak (we measure $-14.5 \pm 0.5$ \AA \ on day 155, with similar values
and larger uncertainty in the lower-resolution spectra around the same
time) to $-92.6 \pm 3$ \AA \ as it was fading quickly on day 197.
H$\alpha$ seen in emission for a longer time may be indicative of a
higher mass-loss rate in NGC~4490-OT than V838~Mon, and in this
respect, NGC~4490-OT is more akin to LBVs with H$\alpha$ emission in
eruption.  Unfortunately, little work has been done using
radiative-transfer models to derive quantitative physical parameters
from the H$\alpha$ strength in a giant LBV eruption.

Even without such radiative-transfer models, however, the H$\alpha$
line profile can provide some interesting physical information about
the kinematics of the NGC~4490-OT transient
event. Figure~\ref{fig:halpha} shows the H$\alpha$ line profile in our
five epochs of spectroscopy for this object.  In this plot, zero velocity
is determined from the centroid of extended narrow (unresolved)
H$\alpha$ emission along the slit, arising from H~{\sc ii} regions in
the same region of the host galaxy.  Of these 5 epochs, the two MMT
spectra with relatively high resolution (1200 line mm$^{-1}$ grating; days 155
and 197) are the most informative.

The H$\alpha$ line near peak (day 155) shows an asymmetric
P~Cygni-like profile.  (The spectra from days 154, 165, and 168 are
consistent with a similar line profile observed with degraded
resolution.)  In Figure~\ref{fig:halpha}, we take the red side of the
line and reflect it about zero velocity and plot it over the
blueshifted side of the line; this reflected red wing is shown in red.
At velocities of $-$650 to $-$1200 km s$^{-1}$, the reflected red wing
overlaps well with the blue wing, but the blueshifted side of
H$\alpha$ shows a pronounced deficit at 0 to $-$650 km s$^{-1}$.  This
comparison gives a strong indication that the underlying emission
component of H$\alpha$ is a symmetric triangular profile, and that the
asymmetry is caused by strong blueshifted self-absorption arising from
a dense outflow with expansion speeds up to 650 km s$^{-1}$.  The blue
edge ($-$650 km s$^{-1}$, representing the maximum outflow speed along
the line of sight) and the centroid of the P-Cygni absorption trough
(roughly $-$280 km s$^{-1}$, perhaps representing the bulk outflow
speed) are marked with dashed vertical purple lines in
Figure~\ref{fig:halpha}.  This outflow speed is reminiscent of the
nebula around $\eta$~Car \citep{smith06}, which has expansion speeds
ranging from about 40 km s$^{-1}$ at the equator to 650 km s$^{-1}$ at
the pole.  Note that the extremes of the line wings are probably
broadened by electron scattering, and do not necessarily reflect the
maximum expansion speed.

These representative speeds in the P-Cygni absorption profile at peak
luminosity (day 155) become quite interesting when we examine the
later spectrum on day 197, during the rapid decline after peak
(Figure~\ref{fig:halpha}).  It also shows an asymmetric profile, but
this time it has excess blueshifted {\it emission} instead of
absorption.  Indeed, the H$\alpha$ line profile has a strong
blueshifted emission bump with a similar centroid at $-$280 km
s$^{-1}$ and a comparable range of speeds as the absorption trough
seen earlier on day 155.  In Figure~\ref{fig:halpha}, we overplot the
day 155 H$\alpha$ profile (in blue) on the day 197 spectrum for
comparison, and the purple vertical dashed lines show speeds of $-$650
and $-$280 km s$^{-1}$ for reference.

The H$\alpha$ profile with the strong blue emission bump on day 197 is
quite unusual, but some similar features have been seen before.  A
comparable blue emission bump in H$\alpha$ was recently reported in
the late-time spectra of UGC~2773-OT, which is a decade-long LBV-like
eruption similar to $\eta$ Carinae \citep{smith16}.  In the same
analysis, \citet{smith16} also showed that line emission from the
present-day bipolar nebula around $\eta$ Car shows a similar blue
emission bump.  An asymmetric blue bump in H$\alpha$ has also been
seen at faster speeds in several SNe~IIn
\citep{smith12,smith12b,smith15b,fransson14}, usually attributed to
bipolar or perhaps disk-like geometry in the shock interaction.  A
corresponding red bump is usually assumed to be weaker or absent owing
to extinction by dust or occultation by the opaque SN ejecta.
Radiative-transfer simulations of SNe~IIn with bipolar CSM support
these expectations of having an asymmetric blue emission line arising
from a bipolar geometry \citep{dessart15}.

A plausible interpretation is that a massive shell was ejected at
these speeds and was initially seen in absorption, but as the
underlying photosphere cooled (as indicated by the spectral evolution
during the decline from peak) and as the optical depth dropped, the
same dense shell is seen in H$\alpha$ emission.  This likely indicates
ongoing shock heating of this ejected shell.  The fact that the line
is asymmetric at late times either requires that the mass ejection was
intrinsically asymmetric with most of the mass ejected toward our
observing direction, or that the receding side of the ejected shell is
obscured by large amounts of dust that formed in the outflow.  The
large IR excess flux and fading of the optical source at late times
are qualitatively consistent with copious dust formation in the
ejecta.

\subsection{Stellar Merger Events Across a Diverse Range of Initial Mass}

In preceding sections, we noted a qualitative similarity between
NGC4490-OT and Galactic transient events that have been interpreted as
stellar mergers: V1309~Sco \citep{tylenda11} and V838 Mon
\citep{bond03,munari02}.  The observed similarities apply to their
somewhat irregular multi-peaked light curves, as well as to the
morphology of their spectra.  The comparison may extend to other
proposed merger objects as well, such as V4332 Sgr in 1994
\citep{martini99}, OGLE 2002-BLG-360 \citep{tylenda13}, M85-2006-OT1
\citep{srk07}, and M31~RV \citep{rich89}, although we did not discuss
these in detail.  CK~Vul also had a multi-peaked light curve
\citep{shara85}.

The progenitor of NGC~4490-OT detected in pre-eruption {\it HST} data
was a more massive star than either V838 Mon or V1309 Sco.  Based on
our favoured value for the local reddening, we inferred that the
progenitor could be quite similar to the lower-luminosity group of
LBVs \citep{smith04}, with an effective initial mass of around 30\,
M$_{\odot}$.  As noted earlier, NGC~4490-OT appears in a relatively
isolated region of its host galaxy, similar to the lower-luminosity
LBVs \citep{st15}.

If indeed NGC~4490-OT was a merger event, then it would appear to
extend a correlation seen in the lower-mass mergers.  Based on a
statistical analysis of published candidate merger events,
\citet{kochanek14} suggested that there is a correlation between the
progenitor's initial mass and the peak luminosity of the transient
resulting from the putative merger.  NGC~4490-OT seems to agree with
this, since it had a more massive progenitor star than either V1309
Sco and V838 Mon, and also had a substantially higher peak luminosity.
The initial mass is uncertain, but for 20--30 M$_{\odot}$, this agrees
reasonably well with the trend reported by \citet{kochanek14}.

We note another possible correlation as well.  In addition to a
correlation between initial mass and transient peak luminosity, we
note that there may be a correlation between both of these and the
duration of the event.  V838 Mon had a longer duration than V1309~Sco,
and NGC~4490-OT had a significantly longer duration than V838~Mon.
There is a simple physical motivation for both of these trends. A
merger event from more massive stars has more kinetic energy and
angular momentum at the time of the merger, perhaps providing more
thermal energy in the merger and a brighter transient event.
Similarly, more massive stars that merge must shed some fraction of
their mass and angular momentum in the event.  If more mass is
ejected, the diffusion time is longer for the ejected envelope mass.

Interestingly, $\eta$ Carinae would seem to extend these same trends,
if it were considered as some sort of binary interaction event.  It is
a much more massive star, and the transient event was somewhat more
luminous at peak and much longer lasting (a decade) than NGC~4490-OT.
In fact, merger models have already been suggested as possible
explanations for $\eta$ Car's 19th century eruption
\citep{kg85,jsg89,pods10}, although these have the obvious difficulty
of explaining why $\eta$ Car also had multiple major eruptions in the
past.  Grazing collisions at periastron have also been discussed as
playing a role in the eruptions \citep{smith11b}.  The emitting
photosphere during $\eta$ Car's eruption was bigger than the current
periastron separation, so some sort of violent interaction must have
occurred \citep{smith11b}, although whether this was a driving
mechanism, a trigger, or merely an after-effect remains uncertain.  It
is interesting that $\eta$ Car shares many of the observational
hallmarks of the other stellar merger events: irregular multi-peaked
light curve \citep{sf11}, transition to redder colours after peak
\citep{prieto14}, followed by dust formation and post-outburst
obscuration \citep{smith03}. We have shown in this paper that
published light-echo spectra of $\eta$~Car \citep{rest12} also
resemble our spectra of NGC~4490-OT and spectra of V838~Mon at some
epochs.

The physical mechanism to power giant LBV eruptions is still not
understood, but if stellar mergers or collisions are viable culprits,
then it may be useful to consider them in context with low-mass
stellar merger events as contributing to the pool of observed non-SN
transient sources.  LBVs may be the more massive extension of these
other eruptive merger events, and that extension may be continuous.
Massive LBV stars are more rare, but in terms of triggering merger
events in close binary systems, LBVs have the physical advantage that
they can change their radius suddenly when they inflate as part of
their S~Doradus variability.  This change may instigate mass transfer
or a merger in a binary system that was previously not interacting.
Folding in selection effects, their higher mass and energy budgets,
with higher peak luminosities and longer durations, may compensate for
their relative rarity, to allow LBV mergers to make a significant
contribution to observed extragalactic SN impostor statistics.  In any
case, if some or all giant LBV eruptions are indeed produced by merger
or collision events, then it becomes more difficult to confidently
classify SN impostors as LBVs, SN2008S-like objects, or something
else.

\section*{Acknowledgements}

\scriptsize 

Some of the data reported here were obtained at the MMT Observatory, a
joint facility of the University of Arizona and the Smithsonian
Institution.  We thank the staffs at Lick and MMT Observatories for
their assistance with the observations. We also appreciate the help of
Jeff Silverman for some of the Lick observations.  Data from Steward
Observatory facilities were obtained as part of the observing program
AZTEC: Arizona Transient Exploration and Characterization.  Lindsey
Kabot assisted with early stages of the MMT spectral data reduction.
The work presented here is based in part on observations made with the
NASA/ESA {\it Hubble Space Telescope}, obtained at the Space Telescope
Science Institute, which is operated by the Association of
Universities for Research in Astronomy, Inc., under NASA contract
NAS5-26555.  These are based in part on observations associated with
program \#13364 (Legacy ExtraGalactic UV Survey, LEGUS). This paper
has made use of the higher-level data products provided by the LEGUS
team.
%

%
%

N.S.\ and J.E.A.\ received partial support from National Science
Foundation (NSF) grants AST-1210599 and AST-1312221.
M.M.K.\ acknowledges support from the Carnegie-Princeton fellowship.
Funding for this effort was provided in part by the {\it Spitzer}
SPIRITS Cycles 10--12 exploration science program.  The supernova
research of A.V.F.'s group at U.C.\ Berkeley presented here is
supported by Gary \& Cynthia Bengier, the Christopher R. Redlich Fund,
the TABASGO Foundation, and NSF grant AST-1211916.  KAIT and its
ongoing operation were made possible by donations from Sun
Microsystems, Inc., the Hewlett-Packard Company, AutoScope
Corporation, Lick Observatory, the NSF, the University of California,
the Sylvia \& Jim Katzman Foundation, and the TABASGO Foundation.
Research at Lick Observatory is partially supported by a generous gift
from Google.  J.J.\ is supported by an NSF Graduate Research
Fellowship under Grant No. DGE-1144469.

\end{document}